\documentclass{article}
\usepackage{authblk}
\usepackage{setspace}
\usepackage{blindtext}
\usepackage{indentfirst}
\usepackage{amsmath}
\usepackage{amsbsy}
\usepackage[mathscr]{euscript}

\begin{document}

\title{Topological Implications of the Total Generalized Electron-Flow Magnetic Helicity Invariant in Electron Magnetohydrodynamics}
\date{}
\author{B.K. Shivamoggi and M. Michalak \\ University of Central Florida \\ Orlando, Fl, 32816-1364 }
\affil{}
\maketitle

\begin{abstract}
Topological implications of the total generalized electron-flow magnetic helicity $H_e$ in electron magnetohydrodynamics (EMHD) are explored. The invariance of $H_e$ is shown to imply the invariance of the sum of the linkage of magnetic field lines, the linkage of electron-flow vorticity field lines and the mutual linkage among these two sets of field lines. This result appears to support a change in the magnetic field topology and hence pave the way for magnetic reconnection in EMHD via a change in the concomitant electron-flow vorticity topology.
\end{abstract}

\section{Introduction}

Topological aspects have been intensely explored in fluid and plasma problems only recently (Moffatt and Tsinober \cite{MT}, Arnol'd and Kesin \cite{AK1}, \cite{AK2}). One measure associated with topological aspects for the latter class of problems is provided by the helicity integral (Woltjer \cite{Woltjer}, Moreau \cite{Moreau}, Moffatt \cite{Moffatt}),

\begin{equation}
H = \int_V \textbf{A} \cdot \textbf{B} d\textbf{x}
\label{1.1}
\end{equation}
where $V$ is the volume occupied by the plasma, which is enclosed by a perfectly conducting magnetic surface $S$, on which $\hat{\textbf{n}}\cdot \textbf{B}=0$. As the curl of a vector field measures its rotation around a point, the integrand in (\ref{1.1}) describes how much the magnetic vector potential, \textbf{A}, rotates around itself. So, $H$ gauges the extent to which the magnetic field lines wind around one another and hence resemble helices. Further, the helicity integral is gauge invariant because it will not change under the gauge transformation $\textbf{A} \Rightarrow \textbf{A}+\nabla \phi$ (which leaves \textbf{B} unchanged).\footnote{This property underscores the physical significance of $H$ because the gauge has no physical significance, so the choice of gauge should not affect \textbf{B}.}

H also provides a topological measure of the structure of the magnetic field because it is directly related to the most basic topological invariant of two linked curves,  1 and 2 (Moffatt \cite{Moffatt}) - their linking number $L_{1,2}$, which was originally given by Gauss in 1833.

This may be quickly seen by noting, subject to the Coulomb gauge $\nabla \cdot \textbf{A}=0$, that

\begin{equation}
\textbf{A}(\textbf{x}) = \frac{1}{c} \int \frac{\textbf{J}(\textbf{x}\prime)}{\mid \textbf{x}- \textbf{x}\prime \mid }d\textbf{x}\prime
\label{1.2}
\end{equation}
Using Ampere's Law,

\begin{equation}
\nabla \times \textbf{B} = \frac{1}{c} \textbf{J}
\label{1.3}
\end{equation}
(\ref{1.2}) becomes

\begin{equation}
\textbf{A}(\textbf{x})=-\int \frac{(\textbf{x} -\textbf{x}\prime)}{\mid \textbf{x}- \textbf{x}\prime \mid^3}\times \textbf{B}(\textbf{x}\prime) d\textbf{x}\prime
\label{1.4}
\end{equation}
Using (\ref{1.4}),  (\ref{1.1}) becomes

\begin{equation}
H=- \iint \limits_V \textbf{B}(\textbf{x}) \cdot \frac{(\textbf{x} -\textbf{x}\prime)}{\mid \textbf{x}- \textbf{x}\prime \mid^3}\times \textbf{B}(\textbf{x}\prime) d\textbf{x} d\textbf{x}\prime
\label{1.5}
\end{equation}
Comparison of (\ref{1.5}) with the linking number\footnote{The linking number describes the linking of two curves and represents the number of times each curve winds around the other.} of the two closed lines,

\begin{equation}
L_{1,2}= - \oint \limits_1 \oint \limits_2 \frac{d \textbf{x}}{d \sigma} \cdot \frac{(\textbf{x}-\textbf{y})}{\mid \textbf{x}-\textbf{y} \mid^3} \times \frac{d \textbf{y}}{d \tau} d\tau d\sigma
\label{1.6}
\end{equation}
where $\sigma$ and $\tau$ are parameters characterizing curves 1 and 2, respectively, leads to the interpretation of $H$ as the double sum of linking numbers over all pairs of magnetic field lines (Moffatt \cite{Moffatt}). Hence, $H$ measures the linkage (or degree of knottedness) of magnetic field lines.\footnote{If the magnetic field lines are bundled into magnetic flux tubes, additional contributions to $H$ come from the twist and writhe of the individual flux tubes (Berger and Field \cite{BF}).}

On the other hand, $H$ turns out to be an invariant (a Hopf invariant\footnote{The Hopf invariant is an invariant of a divergence-free vector field on a three-dimensional (3D) manifold under the group of volume preserving diffeomorphisms and describes the helicity of the field.}) for an ideal plasma enclosed by a perfectly conducting magnetic surface (Woltjer \cite{Woltjer}, Moffatt \cite{Moffatt}) which implies the conservation of linkage (or the degree of knottedness) of the magnetic field lines\footnote{The magnetic field lines cannot penetrate through each other in an ideal plasma.} and, hence, the topology of the magnetic field lines in an ideal plasma is fixed.

In the magnetohydrodynamic (MHD) model, the ions dominate the dynamics while the electrons merely serve to shield out any charge imbalances. On the other hand, in the electron MHD (EMHD) model, with length scales $\rho_e \ll \ell \ll \rho_i$, $s = i,e,\rho_s$ is the gyro-radius, $\rho_s \equiv v_{ts}/\omega_{cs}, v_{ts}$ being the thermal spread of the particles and $\omega_{cs}$ being the cyclotron frequency $\omega_{cs}=e_sB/m_s c$, electrons dominate the dynamics while the demagnetized ions merely serve to provide the neutralizing static background (Kingsep et al. \cite{Kingsep}, Gordeev et al. \cite{Gordeev}). The EMHD model restricts itself to length scales $ \ell \ll d_i$, where $d_s$ is the skin depth, $d_s \equiv c/\omega_{ps}$, $\omega_{ps}$ is the plasma frequency, $\omega_{ps} \equiv \sqrt{n_s e_s^2/m_s}$, and frequencies $\omega > \omega_{ci},\omega_{pi}$. The frozen-in condition of the magnetic field in EMHD is destroyed by electron inertia. Observations of plasma in the magnetosphere (Deng and Matsumoto \cite{DengMatsumoto}) and laboratory (Ren et al. \cite{Ren}) showed that the magnetic reconnection process is initiated in very thin current sheets (thickness on the order of $d_e$).

The EMHD equations admit the total generalized electron-flow magnetic helicity invariant $H_e$ (Shivamoggi \cite{Shivamoggi}),

\begin{equation}
H_e = \int \limits_V \textbf{A}_e \cdot \textbf{B}_e d\textbf{x}
\label{1.7}
\end{equation}
where $\textbf{B}_e$ is the generalized magnetic field,

\begin{equation}
\textbf{B}_e \equiv \nabla \times \textbf{A}_e
\label{1.8}
\end{equation}
and $\textbf{A}_e$ is the generalized magnetic vector potential,

\begin{equation}
\textbf{A}_e \equiv \textbf{A}-\frac{m_e c}{e}\textbf{v}_e
\label{1.9}
\end{equation}
and $V$ is the volume occupied by the plasma. The purpose of this paper is to explore the topological implications of (\ref{1.7}) for EMHD.

\section{Total Generalized Electron Magnetic Helicity Invariant}
The EMHD equations are summarized by

\begin{equation}
\frac{\partial \boldsymbol{\Omega}_e}{\partial t} = \nabla \times (\textbf{v}_e \times \boldsymbol{\Omega}_e)
\label{2.1}
\end{equation}
where $\boldsymbol{\Omega}_e$ is the generalized electron-flow vorticity,

\begin{equation}
\boldsymbol{\Omega}_e \equiv \boldsymbol{\omega}_e+\boldsymbol{\omega}_{ce}, \hspace{.5cm} \boldsymbol{\omega}_e \equiv \nabla \times \textbf{v}_e, \hspace{.5cm} \boldsymbol{\omega}_{ce} \equiv -\frac{e \textbf{B}}{m_e c}.
\label{2,2}
\end{equation}

Equation (\ref{1.1}) has the Hamiltonian formulation (Shivamoggi \cite{Shivamoggi}),

\begin{equation}
H=\frac{1}{2} \int \limits_V \boldsymbol{\psi}_e \cdot \boldsymbol{\Omega}_e d\textbf{x}
\label{2.3}
\end{equation}
where

\begin{equation}
m_e n_e \textbf{v}_e \equiv \nabla \times \boldsymbol{\psi}_e
\label{2.4}
\end{equation}
The non-uniqueness implicit in (\ref{2.4}) may be resolved by imposing the gauge condition

\begin{equation}
\nabla \cdot \boldsymbol{\psi}_e=0
\label{2.5}
\end{equation}
Using (\ref{2.4}), (\ref{2.3}) leads to the expected result,

\begin{equation}
H= \frac{1}{2}\int \limits_V (m_e n_e \textbf{v}_e^2+\textbf{B}^2) d\textbf{x}.
\label{2.6}
\end{equation}

Choosing $\boldsymbol{\Omega}_e$ to be the canonical variable, and following Olver \cite{Olver}, we take

\begin{equation}
\textbf{J}=-\nabla \times\Big[ \Big(\frac{\boldsymbol{\Omega}_e}{m_en_e}\Big)\times(\nabla \times (\bullet))\Big]
\label{2.7}
\end{equation}
as an $\boldsymbol{\Omega}_e$-depedent differential operator which produces a skew-symmetric transformation of the vector function vanishing on the surface $S$ enclosing $V$.

Hamilton's equation is then,

\begin{equation}
\frac{\partial \boldsymbol{\Omega}_e}{\partial t} = J \frac{\delta H}{\delta \boldsymbol{\Omega}_e}
\label{2.8}
\end{equation}
or

\begin{equation}
\frac{\partial \boldsymbol{\Omega}_e}{\partial t}=-\nabla \times\Big[ \Big(\frac{\boldsymbol{\Omega}_e}{m_en_e}\Big)\times(\nabla \times \boldsymbol{\psi}_e) \Big] = \nabla \times (\textbf{v}_e \times \boldsymbol{\Omega}_e)
\label{2.9}
\end{equation}
as required. Here, $\delta H/\delta q$ is the variational derivative. The Hamiltonian formulation therefore implies that the EMHD equation (\ref{2.1}) characterizes a geodesic flow on an infinite-dimensional group of volume-preserving diffeomorphisms.

The operator $J$ may be seen to induce a Poisson bracket (Kuznetsov and Mikhailov \cite{KM}),

\begin{equation}
\begin{gathered}
\big[F,G\big]=\Big(\frac{\delta F}{\delta \boldsymbol{\Omega}_e}, J \frac{\delta H}{\delta \boldsymbol{\Omega}_e}\Big) \\
=-\int \limits_V \frac{\delta F}{\delta \boldsymbol{\Omega}_e} \cdot \nabla \times \Big[ \Bigg( \frac{\boldsymbol{\Omega}_e}{m_e n_e}\Bigg) \times \Big( \nabla \times \frac{\delta G}{\delta \boldsymbol{\Omega}_e} \Big) \Big]d \textbf{x} \\
=- \int \limits_V \Big( \nabla \times \frac{\delta F}{\delta \boldsymbol{\Omega}_e} \Big) \cdot \Big[ \Bigg(\frac{\boldsymbol{\Omega}_e}{m_e n_e}\Bigg) \times \Big( \nabla \times \frac{\delta G}{\delta \boldsymbol{\Omega}_e} \Big) \Big] d \textbf{x} \\
=-\int \limits_V \Bigg(\frac{\boldsymbol{\Omega}_e}{m_e n_e}\Bigg) \cdot \Big[\Big( \nabla \times \frac{\delta F}{\delta \boldsymbol{\Omega}_e} \Big)\times \Big( \nabla \times \frac{\delta G}{\delta \boldsymbol{\Omega}_e} \Big) \Big] d \textbf{x}
\end{gathered}
\label{2.10}
\end{equation}
which is a bilinear function defined on admissible functionals $F[\boldsymbol{\Omega}_e]$ and $G[\boldsymbol{\Omega}_e]$ satisfying

\begin{equation}
\nabla \cdot \begin{pmatrix}
					\displaystyle\frac{\delta F}{\delta \boldsymbol{\Omega}_e} \\[6pt]
					\displaystyle\frac{\delta G}{\delta \boldsymbol{\Omega}_e}
					\end{pmatrix}
=\textbf{0}
\label{2.11}
\end{equation}
in $V$, and

\begin{equation}
\Biggl| \frac{\delta F}{\delta \boldsymbol{\Omega}_e} \Biggr|, \Biggl| \frac{\delta G}{\delta \boldsymbol{\Omega}_e}\Biggr|=0
\label{2.12}
\end{equation}
on $S$.

The Casimir invariant for this problem, which is the annihilator (with respect to any pairing functional) of the Poisson bracket is, therefore, a solution of the equation

\begin{equation}
J \frac{\delta \mathscr{C}}{\delta \boldsymbol{\Omega}_e} = - \nabla \times \Bigg[ \Big(\frac{\boldsymbol{\Omega}_e}{m_e n_e} \Big) \times \Big( \nabla \times \frac{\delta \mathscr{C}}{\delta \boldsymbol{\Omega}_e} \Big) \Bigg] = \textbf{0}
\label{2.13}
\end{equation}
from which,

\begin{equation}
\frac{\delta \mathscr{C}}{\delta \boldsymbol{\Omega}_e} \equiv -\textbf{A}_e = \textbf{v}_e-\frac{e \textbf{A}}{m_e c}
\label{2.13}
\end{equation}
so,

\begin{equation}
\mathscr{C} = -\int \limits_V \textbf{A}_e \cdot \boldsymbol{\Omega}_e d\textbf{x} = C \int \limits_V \textbf{A}_e \cdot \textbf{B}_e d\textbf{x} = C H_e
\label{2.14}
\end{equation}
which is the total generalized electron-flow magnetic helicity $H_e$, $C$ being a constant. The invariance of $H_e$ may be expected to signify some restrictions on the topological aspects of the magnetic field and electron vorticity in EMHD. We will now explore the topological implications of the invariant $H_e$ for EMHD.

\section{Topological Implications of the Total Generalized Electron-Flow Magnetic Helicity Invariant}
On noting that the generalized magnetic flux function $\textbf{A}_e$ satisfies the Coulomb gauge

\begin{equation}
\nabla \cdot \textbf{A}_e = 0
\label{3.1}
\end{equation}
we have,

\begin{equation}
\textbf{A}_e(\textbf{x}) = \int \limits_V \frac{\nabla' \times \textbf{B}_e(\textbf{x}\prime)}{\mid \textbf{x}-\textbf{x}\prime \mid} d\textbf{x}\prime
\label{3.2}
\end{equation}
or

\begin{equation}
\textbf{A}_e(\textbf{x}) = - \int \limits_V \frac{(\textbf{x}-\textbf{x}\prime)}{\mid \textbf{x}-\textbf{x}\prime \mid^3} \times \textbf{B}_e(\textbf{x}\prime) d\textbf{x}\prime.
\label{3.3}
\end{equation}
Using (\ref{3.3}), (\ref{2.14}) becomes

\begin{equation}
H_e = - \iint \limits_V \textbf{B}_e(\textbf{x}) \cdot \frac{(\textbf{x}-\textbf{x}\prime)}{\mid \textbf{x}-\textbf{x}\prime \mid^3}\times \textbf{B}_e(\textbf{x}\prime) d\textbf{x} d\textbf{x}\prime
\label{3.4}
\end{equation}
where $C$ is taken to be unity. Using (\ref{1.9}) (or (\ref{2,2})), (\ref{3.4}) becomes

\begin{equation}
\begin{gathered}
H_e = \iint \limits_V (\textbf{B}(\textbf{x})-d_e^2 \nabla^2 \textbf{B}(\textbf{x}))\times (\textbf{B}(\textbf{x}\prime)-d_e^2 \nabla'^2 \textbf{B}(\textbf{x}\prime)) \\
\cdot \frac{(\textbf{x}-\textbf{x}\prime)}{\mid \textbf{x}-\textbf{x}\prime \mid^3} d\textbf{x} d\textbf{x}\prime
\end{gathered}
\label{4.1}
\end{equation}
or

\begin{equation}
\begin{gathered}
H_e = \iint \limits_V   \textbf{B}(\textbf{x})\times \textbf{B}(\textbf{x}\prime) \cdot \frac{(\textbf{x}-\textbf{x}\prime)}{\mid \textbf{x}-\textbf{x}\prime \mid^3} d\textbf{x} d\textbf{x}\prime \\
-d_e^2 \iint \limits_V \textbf{B}(\textbf{x})\times \nabla'^2 \textbf{B}(\textbf{x}\prime) \cdot \frac{(\textbf{x}-\textbf{x}\prime)}{\mid \textbf{x}-\textbf{x}\prime \mid^3} d\textbf{x} d\textbf{x}\prime \\
- d_e^2 \iint \limits_V  \nabla^2 \textbf{B}(\textbf{x})\times \textbf{B}(\textbf{x}\prime) \cdot \frac{(\textbf{x}-\textbf{x}\prime)}{\mid \textbf{x}-\textbf{x}\prime \mid^3} d\textbf{x} d\textbf{x}\prime \\
+d_e^4 \iint \limits_V  \nabla^2 \textbf{B}(\textbf{x})\times \nabla'^2\textbf{B}(\textbf{x}\prime) \cdot \frac{(\textbf{x}-\textbf{x}\prime)}{\mid \textbf{x}-\textbf{x}\prime \mid^3} d\textbf{x} d\textbf{x}\prime
\end{gathered}
\label{4.2}
\end{equation}
which may be rewritten as,

\begin{equation}
\begin{gathered}
H_e = \iint \limits_V   \textbf{B}(\textbf{x})\times \textbf{B}(\textbf{x}\prime)\cdot \frac{(\textbf{x}-\textbf{x}\prime)}{\mid \textbf{x}-\textbf{x}\prime \mid^3} d\textbf{x} d\textbf{x}\prime \\
+d_e^2 \iint \limits_V \textbf{B}(\textbf{x})\times \big[ \nabla' \times (\nabla' \times \textbf{B}(\textbf{x}\prime))\big]\cdot \frac{(\textbf{x}-\textbf{x}\prime)}{\mid \textbf{x}-\textbf{x}\prime \mid^3} d\textbf{x} d\textbf{x}\prime \\
+d_e^2 \iint \limits_V \big[ \nabla \times(\nabla \times \textbf{B}(\textbf{x}))\big]\times \textbf{B}(\textbf{x}\prime) \cdot \frac{(\textbf{x}-\textbf{x}\prime)}{\mid \textbf{x}-\textbf{x}\prime \mid^3} d\textbf{x} d\textbf{x}\prime \\
+d_e^4 \iint \limits_V  \big[ \nabla \times(\nabla \times \textbf{B}(\textbf{x}))\big]\times \big[ \nabla' \times (\nabla' \times \textbf{B}(\textbf{x}\prime))\big] \cdot \frac{(\textbf{x}-\textbf{x}\prime)}{\mid \textbf{x}-\textbf{x}\prime \mid^3} d\textbf{x} d\textbf{x}\prime 
\end{gathered}
\label{4.3}
\end{equation}
or

\begin{equation}
\begin{gathered}
H_e = \iint \limits_V   \textbf{B}(\textbf{x})\times \textbf{B}(\textbf{x}\prime)\cdot \frac{(\textbf{x}-\textbf{x}\prime)}{\mid \textbf{x}-\textbf{x}\prime \mid^3} d\textbf{x} d\textbf{x}\prime \\
+\frac{m_e^2 c^2}{e^2} \iint \limits_V \boldsymbol{\omega}_e(\textbf{x}) \times \boldsymbol{\omega}_e(\textbf{x}\prime) \cdot \frac{(\textbf{x}-\textbf{x}\prime)}{\mid \textbf{x}-\textbf{x}\prime \mid^3} d\textbf{x} d\textbf{x}\prime\\
-2 \frac{m_e c}{e} \iint \limits_V \textbf{B}(\textbf{x})\times \boldsymbol{\omega}_e(\textbf{x}\prime) \cdot \frac{(\textbf{x}-\textbf{x}\prime)}{\mid \textbf{x}-\textbf{x}\prime \mid^3} d\textbf{x} d\textbf{x}\prime\\
\end{gathered}
\label{4.4}
\end{equation}

Comparison with (\ref{1.6}) shows that the first integral in (\ref{4.4}) gives $L_{B_{1,2}}$, the double sum of the linking numbers over all pairs of magnetic field lines while the second integral gives $L_{\omega_{e_{1,2}}}$ that over all pairs of electron-flow vortex lines. The third integral gives $L_{B_{1},\omega_{e_2}}$ which is the double sum of linking numbers over all pairs of magnetic field lines and electron-flow vortex lines. So (\ref{4.4}) implies

\begin{equation}
L_{B_{1,2}}+K^2 \frac{m_e^2 c^2}{e^2} L_{\omega_{e_{1,2}}}+K \frac{m_e c}{e} L_{B_{1},\omega_{e_2}} = const.
\label{4.5}
\end{equation}
where the constant $K$ is proportional to the ratio of the total electron-flow vorticity flux linkage to the total magnetic flux linkage. So, the invariance of $H_e$ implies the invariance of the sum of the self-linkage of magnetic field lines, the self-linkage of of electron-flow vorticity field lines and the mutual linkage (or knottedness)\footnote{This mutual linkage is similar to that assiciated with cross helicity in MHD (which is an invariant by itself).} amongst these two sets of field lines. This seems to support a change in magnetic field topology and hence pave the way for magnetic reconnection in EMHD via a change in the concomitant electron-flow vorticity topology.

\section{Discussion}
In this paper we have explored the topological implications of the total generalized electron-flow magnetic helicity invariant $H_e$ in EMHD. The invariance of $H_e$ is shown to imply the invariance of the sum of the self-linkage of magnetic field lines, the self-linkage of electron-flow vorticity field lines and the mutual linkage among these two sets of field lines. This result appears to support a change in the magnetic field topology and hence pave the way for magnetic reconnection in EMHD via a change in the concomitant electron-flow vorticity topology. A related result is the reduction of the lower bound on the magnetic energy in EMHD (Shivamoggi \cite{Shivamoggi2}) by an amount proportional to the sum of total electron-flow kinetic energy and total electron-flow enstrophy. This lower bound is produced by the topological barrier provided by the linkage of generalized magnetic field lines underscored by $H_e \neq 0$.

\end{document}